# Solving peak overlaps for proximity histogram analysis of complex interfaces for atom probe tomography data

Jens Keutgen[1], Andrew J. London[2], Oana Cojocaru-Miredin[1,*]

[1] RWTH Aachen, I. Physikalisches Institut (1A), Aachen, Germany

[2] UK Atomic Energy Authority, Culham Science Centre, Oxfordshire OX14 3DB, UK

*corresponding author: cojocaru-miredin@physik.rwth-aachen.de

## Abstract

Atom probe tomography is a powerful tools in investigating nanostructures such as interfaces and nanoparticles in material science. Advanced analysis tools are particularly useful for analyzing these nanostructures characterized very often by curved shapes. However, these tools are very limited for complex materials with non-negligible peak overlaps in their respective mass-to-charge ratio spectra. Usually, an analyst solves peak overlaps in the bulk regions but the behavior at interfaces is rarely considered. Therefore, in this work we demonstrate how the proximity histogram generated for a specific interface can be corrected by using the natural abundances of isotopes. This leads to overlap-solved proximity histograms with a resolution of up to 0.1 nm. This work expands on previous work that showed the advantage of a maximum likelihood peak overlap solving. The corrected proximity histograms together with the maximum likelihood peak overlap algorithm were implemented in an user-friendly software suite called EPOSA.

Keywords: Atom probe tomography, interfaces,

# Introduction

Atom probe tomography (APT) is a nano-analytical characterization technique that uses time-of-flight mass spectrometry to identify individual isotopes in the mass-to-charge ratio spectrum (Kelly & Larson, 2012; Müller et al., 1968). Yet, it is sometimes difficult to distinguish certain ionic isotopes in the mass spectrum as they can overlap with other ionic isotopes due to the limited time-of-flight resolution of the instrument (Haley et al., 2015). To obtain accurate elemental compositions, it is essential that each detected ionic isotope and its corresponding time-of-flight peak is assigned correctly and does not overlap with an isotope of a different ion. Unfortunately, the chances of an overlap between different ions increases at interfaces as ions of both sides are present in this region. This can be a serious problem when investigating numerous materials (Takahashi et al., 2011; Soni et al., 2019), (Blum et al., 2017). If there are relevant overlaps in the material or at an interface, it is necessary to solve them to obtain an accurate composition measurement of the interface region. However, the available software solutions do not offer this possibility, although it is prevalent in many complex materials, ranging from metal alloys to semiconductor compounds (Soni et al., 2015; Narayan et al., 2012; Mancini et al., 2014).

While in the future, kinect-energy information can aid in ion discrimination (Kelly, 2011), currently only the natural abundances of different ions can be used to estimate the overlapped ion amounts in a complex mass spectrum. This was already explored in the past by minimizing the squared residuals (Lawson et al., 1995; Johnson et al., 2013). In a previous work by London et al. (London et al., 2017) a localized peak overlap solution was presented that was able to obtain results with a higher precision by using a maximum likelihood approach to better assign the field-evaporated ions.

Therefore, in this work we show a method, that can successfully generate a corrected proximity histogram, which we call an overlap-solved proximity histogram (OSPH). Using the OSPH we could achieve, for the first time, a very reliable composition investigation of the interface region for the studied materials. More precisely, the creation of isosurfaces maximizes the available data points for the overlap-solving and thus reduces the uncertainties in these evaluations. Additionally, a numerical uncertainty estimation is established to strengthen the analysis of these complex materials.

## Materials and Methods

In this work, two materials and one simulated dataset were chosen as proof-of-principle studies.

First, a simulated dataset was analyzed to investigate the precision of both the isosurface generation and deconvolution. The simulated dataset is a simplified version of a maraging steel with spherical NiAl nanoprecipitates (Ni45-Al45-Fe10) in a FeCr (Fe84-Cr10-Ni3-Al3) matrix that contains Al (3 at. %) and Ni (3 at. %) (Sun et al., 2018). The simulated particles have a radius of 2 nm with a 0.2 nm Gaussian blur applied in x and y direction and 0.1 nm in z direction to mimic the spatial resolution of a real dataset. The simulation box is a cube, size 35 nm, with a mean atomic density of 35 ions/nm$^3$. Simulated data was produced using posgen (London, n.d.), a link to the scripts used can be found in the supplemental material. An overlap diagram for this data set is given in Figure 1, this diagram shows the overlapping ions (bubbles) and which mass positions overlap (labelled lines). The most important overlap in the dataset is between Cr$^{++}$, Fe$^{++}$ and Al$^+$ at 27 Da, influencing the measured content of all Cr, Fe and Al and strongly altering the ratio between Al and Fe. The small particle size combined with the overlaps of multiple isotopes at the same position in the mass spectrum makes it an extreme case, as the limited number of ions in the small particles influences the isosurface creation and the peak overlap-solving. This allows us to explore the limits of the developed overlap-solving algorithm.

Second, we have studied the interface between the n-type Zn(O,S) and the p-type Cu(In,Ga)Se$_2$ absorber layer, which constitutes the p-n junction in the Cu(In,Ga)Se$_2$ thin-film solar cells. Zn(O,S) is an important alternative buffer layer to the existing toxic CdS and a detailed understanding of the interface between this buffer and the absorber is essential when wanting to design more efficient Cu(In,Ga)Se$_2$ thin-film solar cells (Cojocaru-Mirdin et al., 2011). With multiple strong peak overlaps between O, S and Se, as partly shown in Figure 2, along with further overlaps visualised in Figure S.1 and Figure S.2, this interface cannot be correctly analyzed without solving peak overlaps. The selected dataset was acquired by a LEAP 4000 X Si (Cameca, WI, USA) in laser pulsing mode with a laser energy of 5 pJ (50 K base temperature, 250 kHz pulse, 0.5 % detection rate, 355 nm laser). The labelled mass spectrum is shown in Figure S.1 and S.2 of the Supplemental material.

Finally, an oxide dispersion-strengthened steel was investigated, with nominal composition Fe-16Cr-4Al-2W-1Zr-0.5Ti-0.4Y$_2$O$_3$ wt% was prepared by high-energy ball-milling of powder constituents, annealing and subsequent consolidation processing (spark plasma sintering), as presented by W. Li et al (Li et al., 2017). These alloys contain complex oxide phases to strengthen them and a large number of different elements which have an important influence on their performance (London et al., 2015). In the precipitate investigated here, non-negligible peak overlaps between ZrO, Zr, O$_2$ and different peaks containing Cr, Fe and Al were observed which hinders the accurate determination of its composition. Mass spectra are shown in the Supplemental material S.3 and S.4 for the metal-matrix and particle respectively. An overlap diagram showing all the overlapping ions in this system is also shown in the Supplemental material (Figure S.5). The selected dataset was analyzed with a LEAP 3000X HR at 0.4 nJ pulse energy, 200 kHz pulse rate and a 532 nm laser (50 K base temperature).

To generate the OSPH, a 3D reconstruction dataset with the correctly assigned mass-to-charge ratios is necessary to perform the following steps: (1) An interface has to be defined by creating a triangulated surface of constant concentration (isosurface). (2) The minimum distance of each ion from the chosen isosurface is calculated which allows each ion to be assigned to a certain distance bin of the proximity histogram. (3) Each distance bin is treated as an independent dataset and after background correction (provided by AtomProbeLab version 0.1.3), the peak overlap solving is performed using the maximum likelihood method, as explained by London et al. (London, 2019). (4) The composition uncertainty is calculated and the proximity histogram is generated. These steps are described in detail in the following sub-sections.

## 2.1 Isosurface calculation

Before creating an isosurface, the 3D APT dataset is separated into individual cubic voxels with adjustable sizes. Each ion is then assigned to its corresponding voxel with a Gaussian delocalization in place as described in ref. (Larson et al., 2013; Hellman et al., 2003). The delocalization distance is given by half the bin size. The isosurface value is given by the number of selected ions divided by the number of total ions in the voxel. Afterward, the MATLAB isosurface algorithm is used to create the isoconcentration surfaces inside the investigated volume. The isoconcentration surface is calculated using a single ionic identity per mass-peak. By allowing the user to select the size of the voxels, it is possible to control the smoothness of the interface, which can have a significant influence on the quality of the analysis. The dependence of voxel size is shown in the Supplemental material (Figure S.6) and discussed in detail later.

## 2.2 Distance calculation

Each ion of the dataset is assigned to the closest face on the isosurface. Next, the distance from each ion to the closest face on the isosurface is calculated using the method described in (Jones, 1995). Afterwards, the sign is computed Afterwards, the sign is computed using angle weighted pseudo-normals as described in (Baerentzen & Aanaes, 2005). With this information present, a corresponding bin can be assigned to each ion in the dataset.

## 2.3 Separating overlaps

Prior to peak overlap solving, we need first to assign to each mass peak to a potential ion. For that, we use a "range" file which contains the mass window (range) and potential ionic identity of every peak. The peak assignment need not account for overlapped peaks, as long as the mass windows and ions present are correct. EPOSA software (Extended .pos Analysis) can automatically allocate the correct ionic identities to each range. More precisely, if an ion is only present in one range from the range file the software will ensure that all isotopes of this element or molecule are included. This is done by placing range window from

$p_{pos} - w_{min} * \sqrt{p_{pos}}$ to $p_{pos} + w_{max} * \sqrt{p_{pos}}$ in time-of-flight space, where $p_{pos}$ is the peak position and $w$ the width constant (default $w_{min} = 0.015$, $w_{max} = 0.03$). Furthermore, the user may directly specify ions to be included.

The definition and separation of overlaps are discussed in detail by London et al. (London et al., 2017). In summary, the algorithm which solves the overlaps is based on a maximum likelihood estimate of the ionic composition. The assumed natural isotopic abundance [A] is used in combination with the measured counts contained in a set of range windows {r} to produce the following log-likelihood estimate:

$$L(r, pn) = \{r\}^T \log([A]\{p\})$$

An initial guess of the ionic composition {p} is numerically maximized to return the most likely {p} for a given {r} and [A]. Although the maximum likelihood approach is slower than the least-squares method (Michael Maus et al., 2001) for separating overlaps, it is a mathematically more efficient estimator (lower variance) and therefore is superior in the case of limited counts. With minimal time constraints in the analysis of APT data and a maximum calculation time in the region of hours, that is an easy trade-off for obtaining improved results.

## 2.4 Uncertainty estimation

It is indeed difficult to find and define all sources of uncertainties and weight them correctly to obtain a correct estimation of the analysis precision. However, with the counting error being a variance in the number of counts in each peak, it can be easily accounted for. By randomly varying the counts in each peak and performing the same analysis steps explained above, a good estimate of the composition uncertainty can be made for each histogram bin even including the influence of peak overlaps which degrade the precision (London, 2019).

After determining the initial composition for a bin, the number of ions in each range of a bin is drawn from a Poisson distribution (Poisson rate equal to measured peak counts) to reflect the counting statistics of the measurement. The analysis is then repeated for the new peak counts. This process is repeated 300 times and the 95% confidence interval can be estimated from the solutions of the upper 97.5 and lower 2.5 percentiles of the simulated peak count data. This method produces a reliable estimate for the uncertainty arising from peak overlaps in the data.

## 2.5 EPOSA software suite

To generate OSPH, the present algorithm was implemented in EPOSA software, which is able to perform some other extended analyses, as described in refs. (Zhu et al., 2018; Cheng et al., 2019; Rodenkirchen et al., 2020). More specifically, this toolset allows loading of .epos format files and allows the user to perform several complex analyses, like the multi-hit proximity histogram in Figure S.8. For the overlap-solved proximity histogram algorithm presented here, EPOSA gives an easy way of selecting the ions for the isosurfaces and afterward guides the user through all steps in the proximity histogram analysis with convenient export options for the results. A screenshot of the software is shown in the Supplementary information, Figure S.7.

# 3 Results

## 3.1 Simulated dataset

In addition to the two case studies below, the solution was applied to a simulated dataset for verification where the ground truth is known. The results of this example are shown in Figure 3. As seen in Figure 3(a), without applying any overlap-solving, the peak overlap between Al and Fe causes an overestimation of Al in the bulk by about 5 at.%. This is because the 27 Da peak is ranged as Al when in the matrix it is mostly Fe. In the precipitate, without solving the peak overlaps, the composition is roughly correct because the 27 Da peak is primarily Al. Had this peak been identified as Fe, which would have been appropriate for the matrix, then the precipitate would incorrectly appear to contain very little Al.

The composition values obtained from OSPH are shown in Figure 3(b) together with the nominal composition inserted to perform this simulation and the isosurfaces for the creation of the proximity histogram are depicted in Figure 3(c). By comparing them, one can see that the composition was successfully corrected by applying the overlap-solving algorithm described above. The Fe composition was corrected from 78.9 at. % to 85 at. % and the Al composition was corrected from 8 at. % to 3 at.%. After the correction, all four elemental compositions are, within the given uncertainties, in perfect agreement with the simulated ground truth composition. The increasing size of the uncertainty towards the center of the precipitates is due to fewer ion counts with decreasing measured volume. Moreover, the interface width (length from 10 at. % to 90 at. %) on the OSPH is of about 1 nm, which is 0.5 nm bigger than the expected interface width of only 0.5 nm as visible from Figure 3(b). The difference can be explained by the limitations of the isosurface generation and can be partly controlled by the voxel size used for isosurface generation. While bigger voxels will result in smoother isosurfaces, they also increase the interface width of the proximity histogram (see Supplemental material Figure 7). Overall, this analysis shows the general strength and weaknesses of the proximity histogram. While it is possible to separate the ions and get a more detailed analysis than would be possible with a 1D concentration profile, it becomes clear that the generation of isosurfaces itself has a great influence on the results obtained. This is a known problem and has already been discussed in the past (Martin et al., 2016; Barton et al., 2019; Hornbuckle et al., 2015).

## 3.2 Case Study 1: Buffer-absorber interface of thin-film Cu(In,Ga)Se$_2$ solar cells

The investigation of buffer/absorber interfaces in Cu(In,Ga)Se$_2$ solar cells is of high importance for a better understanding and controlling of the cell energy conversion efficiency. For detailed composition measurement, atom probe tomography allows us to detect even small impurities at the interface. Unfortunately, as shown in Figure 2, the Zn(O,S) buffer shows a high number of overlaps in the mass spectrum and, hence, the buffer/absorber interface cannot be analyzed accurately using existing software solutions (Soni et al., 2015). For the example below, the expected approximate atomic composition of the buffer layer is 50% Zn, 30% S and 20% Oxygen.

Figure 4 shows a comparison of the proximity histogram obtained with and without overlap solving. The overlaps between Zn and $S_2$ resulted in the Zn composition inside the buffer to be reduced by almost 10 at. %, while the S composition is overestimated by 5 at. %. Additionally, the uncorrected proximity histogram shows a Se diffusion of about 5 at. % inside the buffer due to overlaps between ZnO/ZnS compounds with Se. This overlap also causes an overestimation of the Zn diffusion inside the absorber layer. While the uncertainties of the Zn and S compositions are larger after overlap solving, this appropriately captures the added uncertainty introduced for the more accurate compositions measurements.

Figure 4 demonstrates the power of the method and visualizes how the improved compositional information allows conclusions which would otherwise not be possible. For example, the details across the interface could not be described by treating the volumes as separate. The improved uncertainty estimation also gives confidence to the obtained results. Additionally, the analysis of separated volumes on either side of the interface is compared with the "Decomposition of Peaks" feature presented in IVAS (Kelly & Larson, 2012), in Table 1. The composition values obtained with the algorithm presented here are in good agreement with the results obtained in IVAS. A comparative study between IVAS and EPOSA could not be done on interfaces since IVAS does not have this capability. This represents also the motivation for the present study.

Table 1: Element composition in the matrix and the buffer. The uncertainties of EPOSA are given as described in section 2.4. The standard deviation of IVAS is the calculated counting errors.

|    | Absorber Composition [at. %] | | | Buffer Composition [at. %] | | |
|----|---|---|---|---|---|---|
|    | IVAS uncorrected | IVAS corrected | EPOSA corrected | IVAS uncorrected | IVAS corrected | EPOSA corrected |
| Cu | 21.7 ± 0.04 | 21.9 ± 0.04 | 21.9 ± 0.03 | 0.3 ± 0.01 | 0.1 ± 0.01 | 0.1 ± 0.01 |
| In | 20 ± 0.04 | 20.3 ± 0.04 | 20.1 ± 0.03 | 0.1 ± 0.01 | 0.1 ± 0.01 | 0.1 ± 0.02 |
| Ga | 9.3 ± 0.03 | 10 ± 0.03 | 9.7 ± 0.02 | 0 | 0 | 0 |
| Se | 46.8 ± 0.05 | 46.4 ± 0.05 | 47.9 ± 0.05 | 2.5 ± 0.02 | 0.7 ± 0.01 | 0.6 ± 0.01 |
| Zn | 1.1 ± 0.01 | 0.6 ± 0.01 | 0.3 ± 0.04 | 37.9 ± 0.04 | 48.1 ± 0.04 | 45.8 ± 0.08 |
| O  | 0.7 ± 0.01 | 0.2 ± 0.01 | 0.1 ± 0.01 | 13.2 ± 0.03 | 15.0 ± 0.03 | 15.5 ± 0.05 |
| S  | 0.1 ± 0.01 | 0.6 ± 0.01 | 0 | 44.7 ± 0.04 | 34.8 ± 0.04 | 36.5 ± 0.1 |
| H  | 0 | 0.02 ± 0.01 | 0 | 1.3 ± 0.01 | 1 ± 0.01 | 1.4 ± 0.02 |

### 3.3 Case Study 2: Oxide nano-precipitates in ODS steel

The final example is of a core-shell oxide precipitate within an ODS steel. In this example, there are different overlapping peaks in the matrix and particle as well as in the Al-oxide shell that surrounds the particle. STEM-EDX of particles from the same material show the matrix has a Fe-15Cr-~5Al composition (at%) composition, the particles have a Y:Zr-rich core with a ratio of about 35:65 surrounded by Al2O3. Figure 5 shows the OSPH and the uncorrected proximity histogram obtained at the interface between the oxide nano-precipitate and the Fe-16Cr matrix. The composition in the matrix of the material (left side) changes slightly due to the Al-Fe-Cr overlap and a more realistic Al content is measured. The Al-content of the Al-oxide shell at the interface remains unchanged, however the influence of the overlap-solved proximity histogram is visible inside the particle itself. The O composition inside the precipitate was corrected by 6 at. % and the Fe composition which was previously overestimated by 5 at. % was also successfully corrected. In addition to this, a 9 at. % composition of Cr was inside the precipitate without any OSPH although the corrected composition shows only 0.1 at. % of Cr inside. Most of the Cr in the particle in the uncorrected composition arises from the ZrO-Cr overlap after overlap solving the Zr and Y content in the particle is 23.0 ± 0.3 and 11.9 ± 0.2, which agrees with the STEM-EDS analysis. Table 2 compares the values obtained by EPOSA for the matrix and precipitate with the bulk (uncorrected) and peak decomposed (corrected) values of IVAS. This table shows that EPOSA and the peak decomposition result from IVAS are consistent. With a total of 11 different elements and 30 ions ranged in this highly complex mass spectrum, the algorithm proved effective at correcting the composition.

Table 2: Element composition in the matrix and the precipitate. The uncertainties of EPOSA are given as described in section 2.4. The standard deviation of IVAS is the calculated counting errors.

|    | Matrix Composition [at. %] | | | Precipitate Composition [at. %] | | |
|----|---|---|---|---|---|---|
|    | IVAS uncorrected | IVAS corrected | EPOSA corrected | IVAS uncorrected | IVAS corrected | EPOSA corrected |
| Fe | 74.5 ± 0.1 | 79.0 ± 0.1 | 79.8 ± 0.1 | 5.97 ± 0.22 | 1.40 ± 0.11 | 0.91 ± 0.11 |
| O  | 0.3 ± 0.02 | 0.16 ± 0.01 | 0.06 ± 0.03 | 55.7 ± 0.5 | 61.5 ± 0.4 | 62.8 ± 0.4 |
| **Y**  | 0.17 ± 0.01 | 0.01 ± 0.01 | 0.01 ± 0.01 | 12.4 ± 0.3 | 11.6 ± 0.3 | 11.9 ± 0.2 |
| **Cr** | 13.2 ± 0.1 | 13.7 ± 0.1 | 13.6 ± 0.1 | 9.35 ± 0.27 | 0.33 ± 0.05 | 0.07 ± 0.04 |
| **Al** | 8.1 ± 0.1 | 3.18 ± 0.06 | 3.17 ± 0.08 | 0.79 ± 0.08 | 0.62 ± 0.07 | 0.69 ± 0.07 |
| **Zr** | 0.47 ± 0.02 | 0.1 ± 0.01 | 0.02 ± 0.02 | 14.0 ± 0.3 | 23.6 ± 0.4 | 23.0 ± 0.3 |

## 4 Discussion

In the examples shown above, the combination of isosurface creation, proximity histogram calculation, and maximum likelihood overlap-solving proved to be a suitable solution for investigations of interfaces in complex mass spectra with multiple overlaps. While a normal per ion overlap-solving can only obtain a resolution in the range of a few nanometers (London et al., 2017), we are able to increase this resolution to up to 0.1 nm, depending on the size of the isosurface. This is possible through the high number of available ions for each bin. Using a larger number of ions for the overlap solving reduces random error and helps avoid bias (London, 2019). As shown in Figure 6, for the large area interface (10,000 nm²) inside the CIGS solar cell, the bin size of 0.1 nm was enough to attain a precision of 48 ± 1 at. % for the Zn composition in the buffer, even with a high number of overlapping peaks and multiple oxide peaks. The simulated dataset shows the benefits of the solution even for many small interfaces. By combining the different interfaces in one analysis, we get enough events in each bin to maintain the proximity histogram resolution of 0.1 nm. The uncertainties in Figure 3 show where the limits of the analysis can be found. The uncertainties increase with decreasing number of ions and are an effective way of verifying the statistical accuracy of the calculation. The comparison of the corrected analysis with the real data also proves that the maximum likelihood approach for the overlap-solving enables detailed and precise results with a highly performant uncertainty estimation. Although this solution is mainly developed for interfaces with many overlaps it could also offer an error correction for other samples, as it uses the additional information of natural abundance in its evaluation of datasets.

The implementation of the presented method into the analysis software EPOSA offers advantages over a function-based implementation in MATLAB. It allows fast and simple creation of isosurfaces on any parameter in the EPOS file and makes the analysis steps easy to follow. Consequently, even users not proficient in programming can install EPOSA and use the proximity histogram with peak overlap-solving. With the focus on data selection and filtering tools, EPOSA and the peak overlap-solving can easily be expanded with additional features in the future.

Besides these benefits of the presented method, it is also important to consider the limitations and potential issues. For obtaining correct results it is indispensable to use a detailed and correct assignment of ions in the corresponding range file. If an ion is not identified in the analysis the influence on the composition of the final result can be drastic. The choice of ranged ions is crucial (London et al., 2017; London, 2019) as the wrong ranging causes a misinterpretation of the abundances in the sample, which can then result in a vastly false composition. Therefore, it is especially important to assign all ranges as accurately and consistent as possible. This includes positioning the ranges correctly and including all present ions. The natural abundance and tools in IVAS or EPOSA for identifying complex peaks can help in finding all present ions, while the range window placing discussed in Section 2.3 can function as guidance for ranging. Inspecting the resulting fit and checking for mis-fitting peaks is also suggested.

The small number of counts of ions in a certain bin also causes difficulties for an accurate analysis. For small numbers of ions, the background correction of the dataset becomes more difficult and the relative peak heights might be influenced. In combination with statistical fluctuations in the number of ions per peak, the natural abundancies may not reflect the detected isotopes and introduce bias to the solution (London, 2019). If the number of events in a bin is insufficient to provide the desired compositional confidence, the width of the histogram bin can easily be increased, thus increasing the number of events per bin but decreasing the spatial resolution of the proximity histogram.

One of the primary sources of uncertainty can come from the isosurface creation itself. While isosurfaces offer a powerful tool to investigate interfaces, they also are not a very precise way of defining these interfaces. The difference between two similar looking isosurfaces can be important for a proximity histogram analysis and can influence the results even in the absence of overlapped ions. Additionally, it is important to keep in mind that the isosurface creation itself has no information on the overlaps inside the sample and assumes a single ionic identity for each mass peak. Hence, it is important to not only perform the analysis for one interface but to verify the results with 1D concentration profiles and optimizing the chosen iso-concentration value, as described in (Yoon et al., 2004).

## Summary

In this work, we have generated peak overlap-solved proximity histograms using the maximum likelihood estimation method. This work allowed us to obtain reliable interface investigations for samples with complex mass spectra with many directly overlapping peaks. The solution was integrated into the easy to use APT data analysis toolset EPOSA. As previously demonstrated, overlap-solving by maximum-likelihood estimation is well suited for small datasets like the individual bins of the calculated proximity histogram. In addition, the integrated uncertainty estimation method strengthens the quality of the results. Hence, the method reported here allows the creation of scientifically relevant analyses of complex interfaces that could previously not be analyzed using the existing tools.

## Acknowledgements

RWTH would like to thank the German Federal Ministry for Economic Affairs and Energy (BMWi) for funding this work within the EFFCIS project under contract number 0324076F.

AJL wishes to thank T. Zhang and the Institute of Solid State Physics, Chinese Academy of Sciences, Hefei, China for providing the ODS material. This work has been carried out within the framework of the EUROfusion consortium and has received funding from the Euratom research and training programme 2014-2018 and 2019-2020 under grant agreement No 633053. The views and opinions expressed herein do not necessarily reflect those of the European Commission.

Figure 1: The overlap diagram of the simulated dataset. The line thickness is proportional to the abundancies of the overlapping isotopes. The most significant overlaps are found at 27 Da between $Fe^{++}$, $Cr^{++}$ and $Al^{+}$.

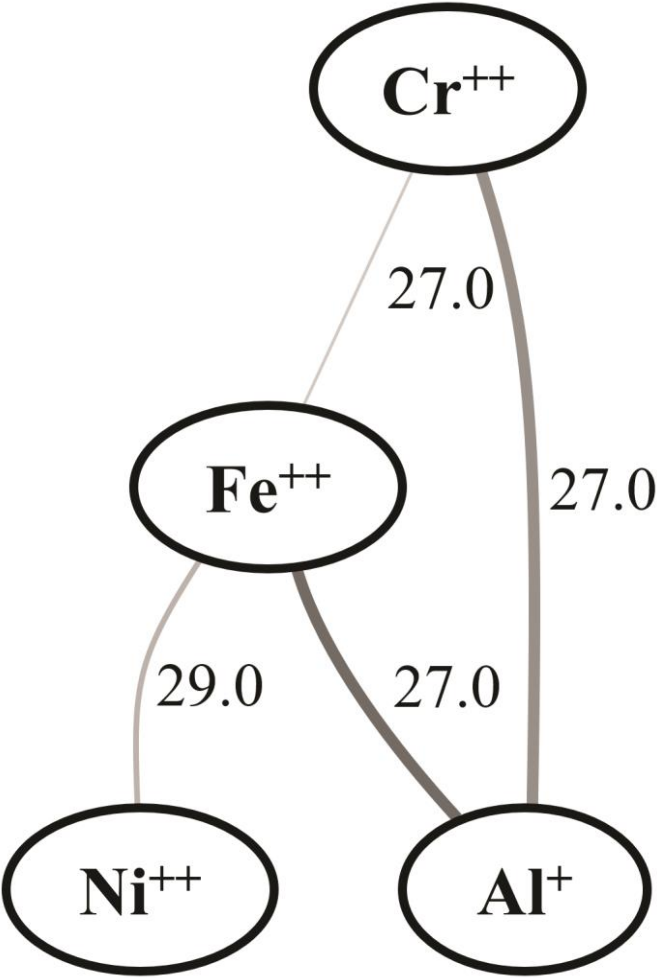

Figure 2: Mass-to-Charge spectrum of the Zn(O,S)/Cu(In,Ga)Se$_2$ interface (red line). Overlaps between $^{64}$Zn$^+$ and $^{64}$S$_2^+$ as well as between $^{66}$Zn$^+$ and $^{66}$S$_2^+$ and $^{65}$Cu$^+$ and $^{65}$S$_2^+$ are detected and influence the results. Labelled lines indicate approximate amounts of different isotopes.

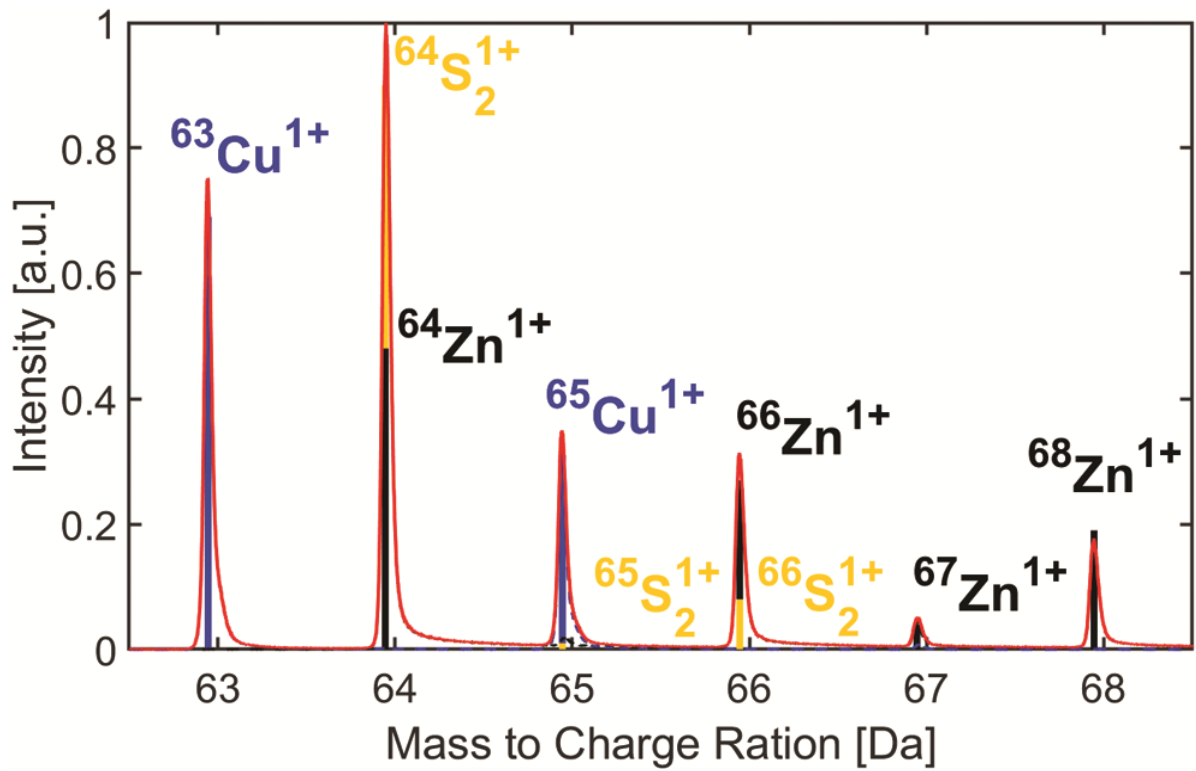

Figure 3: OSPH analysis performed on the simulated dataset. (a) Results without applying peak overlap-solving. The Al composition is overestimated, whereas the Fe composition is underestimated as the 27 Da peak is ranged as Al. (b) The overlap-solved dataset exhibiting the correct composition values. The dashed lines show the true composition of the sample as known from the simulation parameters and the other lines show the corrected dataset. (c) 3D reconstruction of the simulated dataset and 60 at. % Fe-isosurfaces.

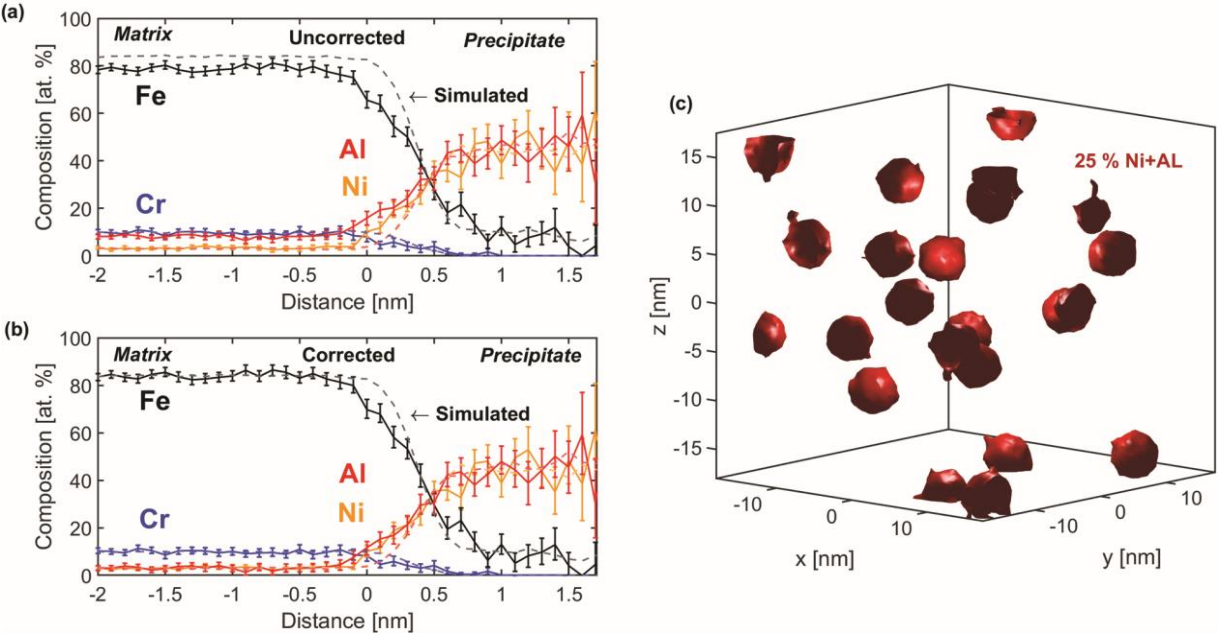

Figure 4: Zn(O,S)/Cu(In,Ga)Se$_2$ Interface without (a) and with (b) peak overlap solving at an isosurface composition of 20% Se (c). (a) In the non-corrected analysis, performed with IVAS, the overlap between Zn and S$_2$ causes a miscalculation of the Zn/S ratio while the overlaps of Se with Zn and S show a strong Se diffusion. (b) The corrected proximity histogram shows an increased Zn composition (47 at. %) while correcting the Se composition (1 at. %) inside the buffer. The Zn diffusion inside the absorber is also corrected from 4 at. % to 1 at. % at a distance of 6 nm.

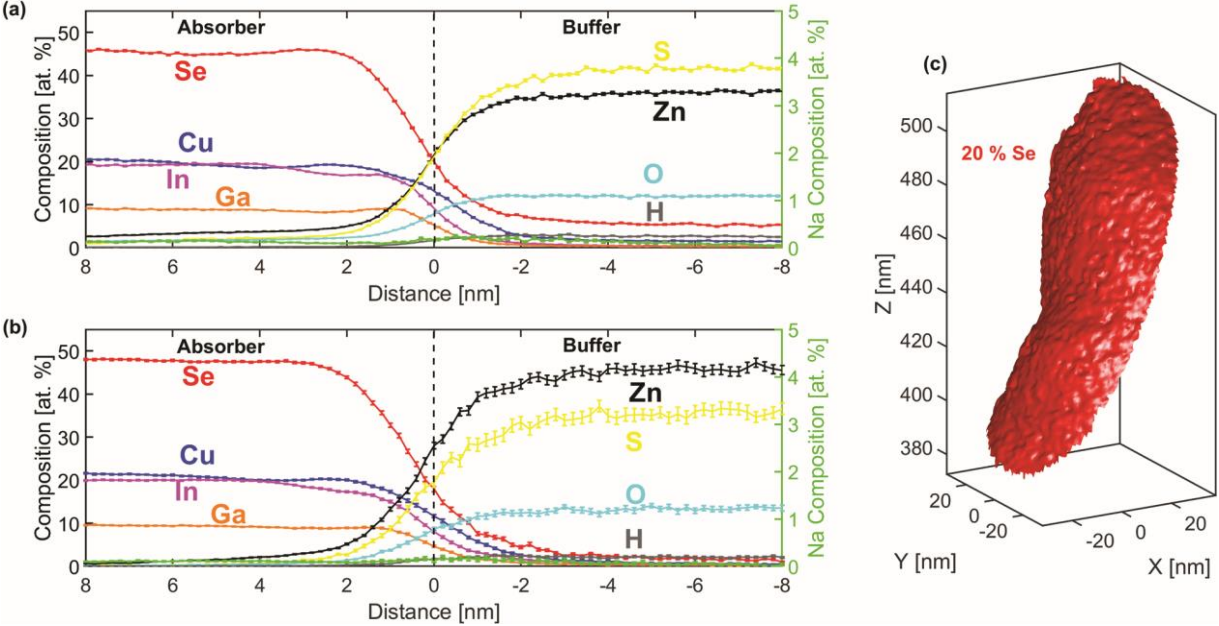

Figure 5: Uncorrected (a) and corrected composition (b) of a proximity histogram along the interface depicted (c). The O content (42 at. %) is underestimated and the Cr (9 at. %) and Fe content (6 at. %) overestimated inside the precipitate without any correction. The corrected proximity histogram shows an O content of (48 at. %) and almost no Fe (2 at. %) or Cr (0.1 at. %) inside the precipitate. Some Ions were left out for clarity. Figure S.9 of the supplementary material shows all ions included in the analysis.

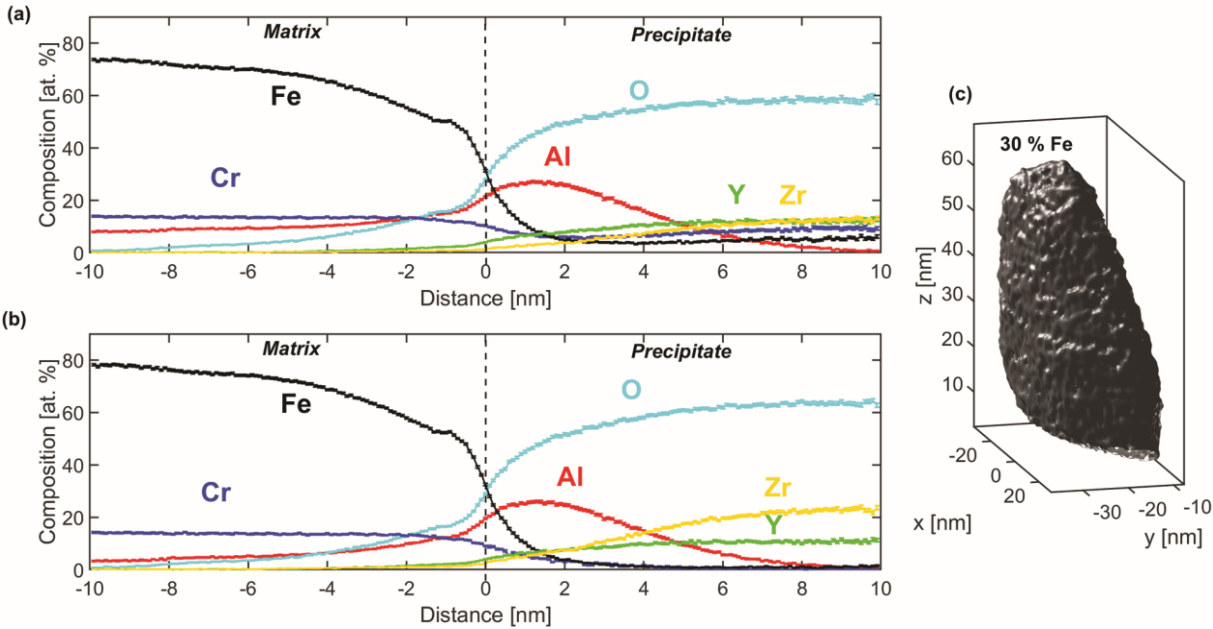

Figure 6: (a) Number of ions for a given range in each bin with a bin size of 0.1 nm. (b) Corresponding mass spectra of the bins at -8 nm and 8 nm. At least $10^2$ ions are available for the analysis in each peak, enabling effective overlap solving even for small bin sizes. For smaller interfaces, an increasing bin size would still enable reliable analysis.

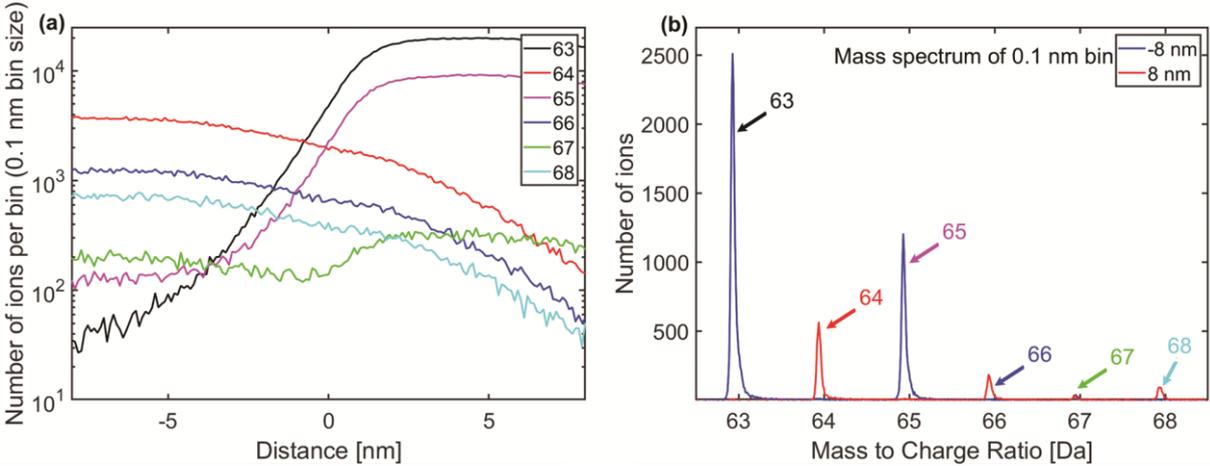